\documentclass[prl,unsortedaddress,superscriptaddress]{revtex4}
\usepackage{amsmath}  
\usepackage{amsbsy,amssymb,amsfonts}
\usepackage{graphicx}
\usepackage{color}
\usepackage{epsf}
\usepackage{wasysym}
\def\ie{i.e.,}
\def\eg{e.g.\ }
\def\cm{$\rm cm^{-1}$}
\def\vac{\left| \rule{0.3cm}{.0cm} \right>}
\newcommand{\mXp}[1]{{\moperator{X}_{#1}^+}}
\def\braket#1{\mathinner{\langle{#1}\rangle}}
\def\Bra#1{\left<#1\right|}
\def\Ket#1{\left|#1\right>}
\def\bravert{\egroup\,\vrule\,\bgroup}
\def\twoint#1{\mathinner{({#1})}}
{\catcode`\|=\active
  \gdef\Twoint#1{\left(\mathcode`\|"8000\let|\bravert {#1}\right)}}
{\catcode`\|=\active
  \gdef\Braket#1{\left<\mathcode`\|"8000\let|\bravert {#1}\right>}}
\newcommand{\mre}[1]{\Re(#1)}
\newcommand{\mim}[1]{\Im(#1)}
\newcommand{\wnb}{cm$^{-1}$}
\newcommand{\mqi}{\check{\imath}}
\newcommand{\mqj}{\check{\jmath}}
\newcommand{\mqk}{\check{k}}
\newcommand{\moperator}[1]{\hat{#1}}
\newcommand{\mqmat}[1]{{{}^Q {#1}}}
\newcommand{\beq}{\begin{equation}}
\newcommand{\eeq}{\end{equation}}
\newcommand{\beqa}{\begin{eqnarray}}
\newcommand{\eeqa}{\end{eqnarray}}
\newcommand{\bea}{\begin{array}}
\newcommand{\eea}{\end{array}}
\newcommand{\bfs}{\bfseries}
\newcommand{\bef}{\begin{figure}}
\newcommand{\ef}{\end{figure}}
\newcommand{\bc}{\begin{center}}
\newcommand{\ec}{\end{center}}
\newcommand{\bt}{\begin{table}}
\newcommand{\et}{\end{table}}
\newcommand{\btb}{\begin{tabular}}
\newcommand{\etb}{\end{tabular}}
\newcommand{\dir}{{\tt {DIRAC}}}
\newcommand{\mea}{{\emph{et al.}}}
\newcommand{\ai}{{\em ab initio}}
\newcommand{\map}{{\emph{a priori}}}
\newcommand{\mapo}{{\emph{a posteriori}}}
\newcommand{\timo}[1]{{\color{magenta}\small \bf Timo: \it #1 }}
\newcommand{\tcr}{\textcolor{red}}
\newcommand{\tcb}{\textcolor{blue}}
\newcommand{\tcg}{\textcolor{green}}
\newcommand{\ket}[1]{\ensuremath{|#1\rangle}\xspace}
\newcommand{\bra}[1]{\ensuremath{\langle #1|}\xspace} 
\def\rvac{\left| \rule{0.3cm}{.0cm} \right>}
\def\lvac{\left< \rule{0.3cm}{.0cm} \right|}
\def\molcas{$\cal M\kern-0.10em O\kern-0.15em L\kern-0.00em 
             C\kern-0.10em A\kern-0.05em S$}
\renewcommand{\thefootnote}{\arabic{footnote}}
\begin{document}

\vspace{2cm}
\title {{ TaO$^+$, a Candidate Molecular Ion in Search of Physics Beyond the Standard Model }}

\vspace*{2cm}

\author{Timo Fleig}
\email{timo.fleig@irsamc.ups-tlse.fr}
\affiliation{Laboratoire de Chimie et Physique Quantiques,
             I.R.S.A.M.C., Universit{\'e} Paul Sabatier Toulouse III,
             118 Route de Narbonne, 
             F-31062 Toulouse, France }

\date{\today}
\vspace*{1cm}
\begin{abstract}
The TaO$^+$ molecular ion is proposed as a candidate system
for detecting signatures of charge parity (${\cal{CP}}$) violating physics beyond the standard
model of elementary particles. The electron electric dipole moment (EDM) effective
electric field $E_{\text{eff}} = 20.2 \left[\frac{\rm GV}{\rm cm}\right]$,
the nucleon-electron scalar-pseudoscalar (ne-SPS) interaction constant $W_{S} = 17.7$ [kHz]
and the nuclear magnetic quadrupole interaction constant 
$W_M = 0.45$ [$\frac{10^{33} {\text{Hz}}}{e\, {\text{cm}}^2}$] 
are found to be sizeable ${\cal{P,T}}$-odd enhancements. The ratio of the leptonic
and semi-leptonic enhancements differs strongly from the one for the ThO system
which may provide improved limits on the electron EDM, $d_e$, and the SPS coupling
constant, $C_S$. TaO$^+$ is found to have
a ${^3\Delta_1}$ electronic ground state like the previously proposed ThF$^+$
molecular ion, but an order of magnitude smaller parallel G-tensor component which
makes it less vulnerable to systematic errors in experiment.
\end{abstract}

\maketitle
\clearpage

\section{Introduction}
\label{SEC:INTRO}

In the search for physics beyond the standard model (BSM) of elementary particles electric dipole moments (EDMs) 
of atomic and molecular systems are a powerful probe in the low-energy regime \cite{ramsey-musolf_review2_2013}. 
The corresponding spatial parity (${\cal{P}}$) and time-reversal (${\cal{T}}$) non-conserving interactions may 
have multiple fundamental sources \cite{khriplovich_lamoreaux,EDMsNP_PospelovRitz2005,ginges_flambaum2004} and are 
themselves, assuming ${\cal{CPT}}$-conservation, manifestations of charge-parity (${\cal{CP}}$) violation beyond 
that already incorporated into the SM \cite{Kobayashi}.

Due to the possibility of having different sources contribute to a ${\cal{P,T}}$-odd effect in an atomic-scale
system, ideally multiple positive measurements on different systems with different dependency on underlying
sources should be sought for \cite{Chupp_Ramsey_Global2015,MJung_robustlimit2013}. One way of making progress
is, therefore, the investigation of new systems that display favorable physical properties in the search for
EDMs.

Measurements and many-body calculations on electronically paramagnetic molecular systems with low-lying 
${^3\Delta_1}$ electronic states currently lead to strong constraints on the electron electric dipole moment, $d_e$, 
and the scalar-pseudoscalar nucleon-electron coupling constant, $C_S$ 
\cite{ACME_ThO_eEDM_science2014,Denis-Fleig_ThO_JCP2016,Skripnikov_ThO_JCP2015,Fleig2014}. In addition, this
kind of paramagnetic molecules also allows for constraining ${\cal{P,T}}$-odd hadron physics, originating in the
quantum chromodynamics (QCD) (${\cal{CP}}$)-violating parameter $\tilde{\Theta}$ \cite{Veneziano_Witten_QCD-PTodd1980},
EDMs of the $u$ and $d$ quarks, $d_{u,d}$, and {\it{via}} chromo-EDMs, $\tilde{d}_{u,d}$ 
\cite{Gunion_Wyler_CEDM_NEDM_1990}. In this context, the interaction of a nuclear magnetic quadrupole moment (MQM) 
with electronic magnetic fields has been addressed for molecules containing strongly deformed thorium and tantalum 
nuclei \cite{fleig:PRA2016,Skripnikov_TaN_PRA2015,Flambaum_DeMille_Kozlov2014,Skripnikov_ThO_PRL2014}.

In the present paper, the TaO$^+$ molecular ion is investigated with four-component relativistic all-electron
methods including the effects of inter-electron correlations and proposed as a promising candidate system for
measurement of an EDM or for further constraining (${\cal{CP}}$)-odd BSM parameters through an experimental null 
result. Molecular properties for low-lying electronic states as well as relevant ${\cal{P,T}}$-odd interaction 
constants for the potential ``science state'', ${^3\Delta_1}$, are presented.

\section{Methods}
\label{SEC:THEORY}
The electron EDM interaction constant is evaluated as proposed in stratagem II of Lindroth et al.
\cite{lindroth_EDMtheory1989} as an effective one-electron operator via the squared electronic
momentum operator,
\begin{equation}
E_{\text{eff}} := \frac{2\imath c}{e\hbar} \left< \Psi_{\Omega} \right|
        \sum\limits_{j=1}^n\, \gamma^0_j \gamma^5_j\, \vec{p}_j\,^2 \left| \Psi_{\Omega} \right>
\end{equation}
with $n$ the number of electrons and $j$ an electron index, as described in greater detail in 
reference \cite{fleig_nayak_eEDM2013}.
The EDM effective electric field is related to the electron EDM interaction constant
$W_d = -\frac{1}{\Omega}\, E_{\text{eff}}$.

The ne-SPS interaction constant is defined and implemented \cite{ThF+_NJP_2015} as
\begin{equation}
 W_{\cal{S}} := \frac{\imath}{\Omega}\,
   \frac{G_F}{\sqrt{2}}\, Z\, \left< \Psi_{\Omega} \right| \sum\limits^n_{j=1}\, {\gamma^0_j\gamma^5_j\, \rho_K(\vec{r}_j)}
   \left| \Psi_{\Omega}  \right>
\end{equation}
where $G_F$ is the Fermi constant, $Z$ is the proton number and $\rho_K(\vec{r}_j)$ is the nuclear charge
density at position $\vec{r}_j$, normalized to unity.

The parallel magnetic hyperfine interaction constant is defined as follows:
\begin{equation}
 A_{||} = \frac{\mu_{K}}{I \Omega}\,
          \left< \Psi_{\Omega} \right| \sum\limits_{i=1}^n\, \left( \frac{\vec{\alpha_i} \times \vec{r}_{iK}}{r_{iK}^3}
          \right)_z \left| \Psi_{\Omega} \right>
\end{equation}
where $\vec{\alpha}$ is a vector of Dirac matrices and $\vec{r}_{iK}$
is the position vector relative to nucleus $K$. Further details can be found in reference \cite{Fleig2014}.

The nuclear MQM interaction constant has been implemented in reference \cite{fleig:PRA2016} and can be written as
\begin{equation}
 W_M = \frac{3}{2\Omega}\, 
 \left< \Psi_{\Omega} \right| 
             -\frac{1}{3}\,   \sum\limits_{j=1}^n\,
			  \left\{ \left[ \alpha_1(j) \frac{\partial}{\partial r_2(j)} -
		                      \alpha_2(j) \frac{\partial}{\partial r_1(j)}
				      \right]\, \frac{r_3(j)}{r^3(j)} \right\}
				      \left| \Psi_{\Omega} \right>.
 \label{EQ:CIEXPECVAL}
\end{equation}
In this case, $r_k(j)$ denotes the $k$-th cartesian component of vector ${\bf{r}}$ for particle $j$ ({\it{idem}} for
$\alpha$).

Finally, the parallel component of the electronic G-tensor for a linear molecule is defined as
\begin{equation}
 \label{EQ:GTENSOR_PAR}
 G_{||} = \frac{1}{\Omega}\, 
          \left< \Psi_{\Omega} | \hat{L}^e_{\hat{n}} + g_s \hat{S}^e_{\hat{n}} | \Psi_{\Omega} \right>
 \end{equation}
for the $N$-electron wavefunction $\Psi_{\Omega}$ in irreducible representation $\Omega$,
with $\hat{L}^e_{\hat{n}} = \hat{\bf{L}}^e \cdot \hat{n}$ where $\hat{n}$ is a unit vector along the
molecular axis and $g_s = -g_e = 2.00231930436182$ is the free-electron g-factor \cite{nist_g-2}.

\section{Results}
\label{SEC:APPL}

\subsection{Technical details}

A local version of the \verb+DIRAC15+ program packages \cite{DIRAC15,knecht_luciparII} has been used for all presented
calculations, extended to allow for the calculation of expectation values over the various reported property operators 
\cite{fleig:PRA2016,fleig_nayak_eEDM2013,Fleig2014,ThF+_NJP_2015}. In all calculations the speed of light was set to 
137.0359998 a.u.

Fully uncontracted all-electron atomic Gaussian basis sets of triple-$\zeta$ quality were used for
the description of electronic shells, in the case of tantalum Dyall's basis set
\cite{dyall_basis_2004,dyall_gomes_basis_2010} and for oxygen the Dunning cc-pVTZ-DK set
\cite{Dunning_jcp_1989}. For tantalum valence- and core-correlating exponents were added,
amounting to \{$30s,24p,15d,11f,3g,1h$\} uncontracted functions.
In electron-correlated calculations the virtual spinor space has been truncated at $30$ a.u.

The two correlated wavefunction models used in the present work, for which the Dirac-Coulomb Hamiltonian 
\begin{equation}
 \label{EQ:HDCMOL}
 \hat{H}^{DC} = \sum\limits_A \sum\limits^N_i\, \left[ c(\vec{{\bf{\alpha}}} 
 \cdot \vec{p})_i + \beta_i m_0c^2 + V_{iA}{1\!\!1}_4 \right]
    + \sum\limits^N_{i,j>i}\, \frac{1}{r_{ij}}{1\!\!1}_4 
    + \sum\limits_{A,B>A}\, V_{AB},
\end{equation}
is diagonalized, are denoted MR$^{+T}_{12}$-CISD($N$), where $\vec{{\bf{\alpha}}}$ is a cartesian vector of Dirac matrices,
$V_{iA}$ is the potential-energy operator for electron $i$ in the electric field of nucleus
$A$, ${1\!\!1}_4$ is a unit $4 \times 4$ matrix and $V_{AB}$ represents the potential energy due to the internuclear
classical electrostatic repulsion of the clamped nuclei, and $N$ is the number of explicitly correlated electrons. 
For the model MR$^{+T}_{12}$-CISD($10$) the model parameter $n=3$ is used in the
present work. Details are to be found in reference \cite{fleig:PRA2016}.

For the determination of the nuclear magnetic hyperfine coupling the tantalum isotope
$\rm {^{181}{T}a}$ is used for which the nuclear magnetic moment is
$\mu = 2.361 \mu_N$ \cite{Ta181_magmom_1973}. Its nuclear spin quantum number is $I = 7/2$.

\subsection{Results and discussion}

\subsubsection{Energetics and spectroscopic properties}

Important characteristics of the active set of molecular four-spinors are displayed in Table \ref{TAB:TAO+:ACTIVESPINORS}.
The large similarities with the isoelectronic TaN system \cite{fleig:PRA2016} are not surprising, but a number of
important differences require special attention.


\begin{table}[h]

\caption{\label{TAB:TAO+:ACTIVESPINORS}
         Characterization of important active-space Kramers pairs in terms of orbital angular momentum projection,
         Dirac-Coulomb Hartree-Fock spinor energy, and principal atomic shell character based on a Mulliken population 
	 analysis at $3.1 a_0$ internuclear distance; 
	 the Kramers pairs numbered 29-35 represent the Ta(4f) shell.
        }

\begin{center}
\begin{tabular}{c|ccrrl}
 No. & \rule{0.3cm}{0.0cm} $|m_j|$ & \rule{0.7cm}{0.0cm} $\left|\left< \hat{\ell}_z \right>_{\varphi_i}\right|$
                    & \rule{0.5cm}{0.0cm} $\varepsilon_{\varphi_i}$ [$E_H$]
                    & \rule{0.5cm}{0.0cm} atomic population, Atom($\ell_{\lambda}$) \\ \hline
 25 &  $1/2$  & $0.000$   & $-3.507$ & \rule{0.5cm}{0.0cm} $99$\% Ta($s$)       \\
 26 &  $1/2$  & $0.592$   & $-2.337$ & \rule{0.5cm}{0.0cm} $38$\% Ta($p_{\sigma}$), $59$\% Ta($p_{\pi}$)        \\
 27 &  $1/2$  & $0.401$   & $-2.028$ & \rule{0.5cm}{0.0cm} $51$\% Ta($p_{\sigma}$), $40$\% Ta($p_{\pi}$)        \\
 28 &  $3/2$  & $1.000$   & $-1.981$ & \rule{0.1cm}{0.0cm} $100$\% Ta($p_{\pi}$)        \\
 $\vdots$ & $\vdots$ & $\vdots$ & $\vdots$ & $\vdots$ \\
 36 &  $1/2$  & $0.006$   & $-1.425$ & \rule{0.1cm}{0.0cm} $84$\% O($s$), $7$\% Ta($p_{\sigma}$), $5$\% Ta($d_{\sigma}$)  \rule{0.0cm}{0.8cm}       \\
 37 &  $1/2$  & $0.021$   & $-0.755$ & \rule{0.5cm}{0.0cm} $70$\% O($p_{\sigma}$), $16$\% Ta($d_{\sigma}$) \\
 38 &  $1/2$  & $0.979$   & $-0.737$ & \rule{0.5cm}{0.0cm} $74$\% O($p_{\pi}$), $20$\% Ta($d_{\pi}$) \\
 39 &  $3/2$  & $1.000$   & $-0.736$ & \rule{0.5cm}{0.0cm} $76$\% O($p_{\pi}$), $20$\% Ta($d_{\pi}$) \\ \hline
 40 &  $1/2$  & $0.001$   & $-0.570$ & \rule{0.5cm}{0.0cm} $79$\% Ta($s$), $17$\% Ta($d_{\sigma}$)       \\
 41 &  $3/2$  & $1.994$   & $-0.556$ & \rule{0.5cm}{0.0cm} $100$\% Ta($d_{\delta}$)        \\
 42 &  $5/2$  & $2.000$   & $-0.555$ & \rule{0.5cm}{0.0cm} $100$\% Ta($d_{\delta}$)        \\ \hline
 43 &  $1/2$  & $0.846$   & $-0.175$ & \rule{0.5cm}{0.0cm} $53$\% Ta($p_{\pi}$), $28$\% Ta($d_{\pi}$), $8$\% Ta($p_{\sigma}$)        \\
 44 &  $3/2$  & $1.003$   & $-0.161$ & \rule{0.5cm}{0.0cm} $63$\% Ta($p_{\pi}$), $33$\% Ta($d_{\pi}$)     \\
 45 &  $1/2$  & $0.154$   & $-0.151$ & \rule{0.5cm}{0.0cm} $44$\% Ta($p_{\pi}$), $26$\% Ta($d_{\sigma}$), $9$\% Ta($d_{\pi}$), $7$\% Ta($d_{\sigma}$)     \\
 46 &  $1/2$  & $0.991$   & $-0.071$ & \rule{0.5cm}{0.0cm} $55$\% Ta($p_{\pi}$), $33$\% Ta($d_{\pi}$), $9$\% O($p_{\pi}$)       \\
 47 &  $3/2$  & $1.003$   & $-0.069$ & \rule{0.5cm}{0.0cm} $54$\% Ta($p_{\pi}$), $35$\% Ta($d_{\pi}$), $8$\% O($p_{\pi}$)  
\end{tabular}

\end{center}
\end{table}


The energy difference $\left| \varepsilon_{\sigma_{6s}}-\varepsilon_{\delta_{5d_{3/2}}} \right| = 0.014\, E_H$ (Kramers
pairs $40$ and $41$) is 
significantly smaller than in TaN ($0.031\, E_H$) and is a consequence of the increased nuclear charge on the neighboring
light atom in TaO$^+$: An electron in a more diffuse Ta-localized $5d_{3/2}$ (or likewise $5d_{5/2}$) state experiences 
a greater stabilisation due the higher mean electrostatic potential than an electron in a more compact Ta-localized 
$6s_{1/2}$ state, leading to a differential energy shift of $-0.017\, E_H \approx -3730$ \cm. This relative
stabilization of the Ta($5d$) spinors in TaO$^+$ is largely responsible for the change of electronic ground state 
compared to TaN, since the relative valence spinor occupation between the two states (${^1\Sigma_0^+}$ and 
${^3\Delta}_1$) predominantly corresponds to a transition $\sigma_{6s(1/2)}^1 \leftrightarrow \delta_{5d(3/2)}^1$.

Also in contrast to TaN (8\% contribution) \cite{fleig:PRA2016} and the corresponding spinor in the YbF molecule (13\%)
\cite{TF_unpub} a $p$-wave contribution to the valence $\sigma_{{6s}_{1/2}}$ spinor is quite small ($\approx 3$\%) in 
the TaO$^+$ system. The fractionally occupied $\sigma_{{6s}_{1/2}}$ spinor is more strongly stabilized by an increase
of nuclear charge on the neighboring atom than the virtual $\sigma_{{6p}_{1/2}}$ spinor (Kramers pair $43$). As a consequence, 
the energy gap is increased and mixing of $s$- and $p$-waves reduced in TaO$^+$.
This fact has implications for the size of ${\cal{P,T}}$-odd matrix elements in the ${^3\Delta_1}$ state and will be 
discussed in a following subsection.

\begin{figure}[h]
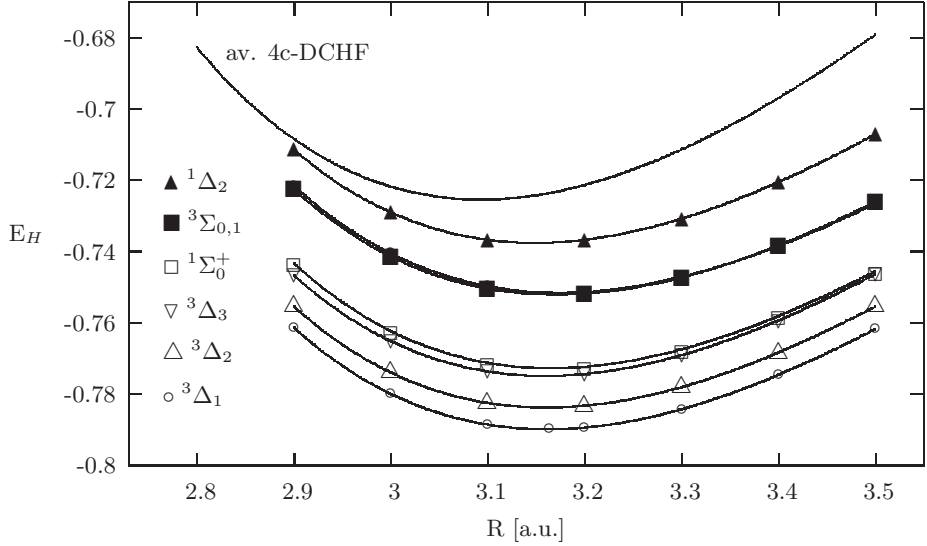

 \caption{\label{FIG:TaN_LOWSTATES}
 Potential energy curves for the lowest-lying electronic state of TaO$^+$, using the CI model MR$^{+T}_{12}$-CISD(10), 
          cutoff $30$ a.u.  The energy offset is $-15671$ E$_H$.
         }

\vspace{0.5cm}

 \begin{center}
\setlength{\unitlength}{0.240900pt}
\ifx\plotpoint\undefined\newsavebox{\plotpoint}\fi
\sbox{\plotpoint}{\rule[-0.200pt]{0.400pt}{0.400pt}}%

 \end{center}
\end{figure}

Important information for experimentally preparing the system in the science state is given by the molecular constants,
transition energies, and transition dipole moments for an ensemble of energetically low-lying electronic states.
The seven lowest electronic states have been addressed in this work. An analysis of the corresponding CI wavefunctions
and results for spectroscopic constants of these states are shown in
Table \ref{TAB:TAN:SPECCON}. Corresponding potential-energy curves are displayed in Fig. \ref{FIG:TaN_LOWSTATES}.


\begin{table}[h]

 \caption{\label{TAB:TAN:SPECCON}
          Characterization of electronic ground and energetically lowest-lying excited states at an internuclear
          distance of $R = 3.1$ a$_0$; the wavefunction model used is MR$^{+T}_{12}$-CISD($10$). The percentage
	  is calculated as $100 \times \sum\limits_j\, \left| C_j \right|^2$ for a normalized CI wavefunction, 
	  where $j$ denotes a Slater determinant and the sum runs over determinants of the respective configuration.
          Spectroscopic constants; $R_e$ the equilibrium internuclear distance, 
	  $\omega_e$ the harmonic vibrational frequency, $B_e$ the rotational constant, 
	  and $T_e$ the equilibrium excitation energy
         }

 \begin{center}
 \begin{tabular}{c|l|cccc}
 ${^{2S+1}\Lambda_{\Omega}}$  &  $\lambda_{n\ell(\omega)_{\text{Atom}}}^o, \omega = |m_j|, o {\text{: occupation}}$
                                    &  $R_e$ [a.u.] &  $\omega_e$ [\cm]  &  $B_e$ [\cm]  &  $T_e$ [\cm]   \\ \hline
 ${^3\Delta_1}$ &  88\% $\sigma_{6s(1/2)_{\text{Ta}}}^1 \delta_{5d(3/2)_{\text{Ta}}}^1$
  &  $3.161$      & $1091$             &  $0.410$      &  $0$           \\
 ${^3\Delta_2}$ &  59\% $\sigma_{6s(1/2)_{\text{Ta}}}^1 \delta_{5d(3/2)_{\text{Ta}}}^1$, 
                                                      29\% $\sigma_{6s(1/2)_{\text{Ta}}}^1 \delta_{5d(5/2)_{\text{Ta}}}^1$
  &  $3.160$      & $1092$             &  $0.410$      &  $1318$        \\
 ${^3\Delta_3}$ &  88\% $\sigma_{6s(1/2)_{\text{Ta}}}^1 \delta_{5d(5/2)_{\text{Ta}}}^1$
  &  $3.160$      & $1093$             &  $0.410$      &  $3270$        \\
 ${^1\Sigma_0^+}$ &  52\% $\sigma_{6s(1/2)_{\text{Ta}}}^2$, 
                                                      32\% $\delta_{5d(3/2)_{\text{Ta}}}^2$, 
                                                       2\% $\delta_{5d(5/2)_{\text{Ta}}}^2$
  &  $3.165$      & $1086$             &  $0.409$      &  $3759$        \\
 ${^3\Sigma_0^+}$ &  12\% $\sigma_{6s(1/2)_{\text{Ta}}}^2$, 
                                                      40\% $\delta_{5d(3/2)_{\text{Ta}}}^2$, 
                                                      35\% $\delta_{5d(5/2)_{\text{Ta}}}^2$
  &  $3.170$      & $1071$             &  $0.408$      &  $8265$        \\
 ${^3\Sigma_1^+}$ &  88\% $\delta_{5d(3/2)_{\text{Ta}}}^1 \delta_{5d(5/2)_{\text{Ta}}}^1$
  &  $3.174$      & $1061$             &  $0.407$      &  $8409$        \\
 ${^1\Delta_2}$ &  27\% $\sigma_{6s(1/2)_{\text{Ta}}}^1 \delta_{5d(3/2)_{\text{Ta}}}^1$, 
                                                      57\% $\sigma_{6s(1/2)_{\text{Ta}}}^1 \delta_{5d(5/2)_{\text{Ta}}}^1$
  &  $3.149$      & $1101$             &  $0.413$      & $11458$
  \end{tabular}
 \end{center}

\end{table}


It is firmly established that ${^3\Delta_1}$ ($\sigma_{6s(1/2)_{\text{Ta}}}^1 \delta_{5d(3/2)_{\text{Ta}}}^1$)
is the electronic ground state of the TaO$^+$ molecular ion. The excitation energy of the ${^1\Sigma_0^+}$ state --
the electronic ground state in the TaN system -- of more than $3700$ [\cm] is much too large to leave doubt about the
principal state ordering. As a further confirmation of the above-discussed effect that leads to this state ordering,
the spin-orbit splitting within the ${^3\Delta}$ multiplet is here about $20$\% larger than in TaN, which is
again due to the increased nuclear charge on the neighboring light atom.

\subsubsection{${\cal{P,T}}$-Odd Properties, Hyperfine Interaction and $G$-Tensor}

Since TaO$^+$ is an electronically paramagnetic system, the dominant sources of a potential ${\cal{P,T}}$-odd effect
are leptonic and semi-leptonic \cite{khriplovich_lamoreaux}. Among these, the pseudoscalar-scalar (PSS) and 
tensor-pseudotensor (TPT) contributions to the molecular EDM are estimated to be at least three orders of magnitude 
smaller than the ne-SPS contribution for many underlying models of (${\cal{CP}}$)-violation 
\cite{Barr_eN-EDM_Atoms_1992}, leaving the electron EDM and ne-SPS interaction as the main contributors.

At the calculated equilibrium internuclear distance and the best level of theory, the effective electric field for 
TaO$^+$ is $E_{\text{eff}} = 20.2 \left[\frac{\rm GV}{\rm cm}\right]$ (see Table \ref{TAB:PTODD_PROP}) which is 
about $45$\% smaller than the value
for TaN \cite{fleig:PRA2016,Skripnikov_TaN_PRA2015}. The reason for this is the smaller $p$-wave contribution
to the valence $\sigma_{{6s}_{1/2}}$ spinor in TaO$^+$, leading to a smaller ${\cal{P}}$-odd matrix element.
$E_{\text{eff}}$ for TaO$^+$ is about $27$\% of the value in the corresponding state for thorium monoxide
\cite{Denis-Fleig_ThO_JCP2016,Skripnikov_ThO_JCP2015} which currently yields the strongest constraints on underlying
BSM parameters \cite{ACME_ThO_eEDM_science2014}.
The ne-SPS interaction constant in Table \ref{TAB:PTODD_PROP} is $W_{S} = 17.7$ [kHz], again about $45$\% smaller 
than the value for TaN which confirms internal consistency of the acquired results.


\begin{table}[h]

\caption{\label{TAB:PTODD_PROP}
         Molecular electric dipole moment, EDM effective electric field,
         magnetic hyperfine interaction constant, scalar-pseudoscalar electron-nucleon interaction constant, 
         nuclear magnetic quadrupole interaction constant and parallel g-tensor component at two internuclear 
         distances $R$ and with two different wavefunction models for the electronic ground state 
	 ${^3\Delta}_1$ ($\Omega = 1$)
        }

\begin{center}
\begin{tabular}{l|ccccccc}
    CI Model, $R$    & $D$ [Debye] & $E_{\text{eff}} \left[\frac{\rm GV}{\rm cm}\right]$ & $A_{||}$ [MHz]
                     & $W_{S}$ [kHz] & $W_{M}$ [$\frac{10^{33} {\text{Hz}}}{e\, {\text{cm}}^2}$] & $G_{||}$ [a.u.] \\ \hline
 MR$^{+T}_{12}$-CISD($10$), $3.1$ a$_0$     & $-3.91$ &  $17.6$  &  $-4537$ &  $15.7$  & $0.38$  & $0.0024$     \\
 MR$^{+T}_{12}$-CISD($18$), $3.1$ a$_0$     & $-3.85$ &  $20.7$  &  $-4593$ &  $18.4$  & $0.46$  & $0.0025$     \\
 MR$^{+T}_{12}$-CISD($10$), $3.1609$ a$_0$  & $-4.08$ &  $17.0$  &  $-4492$ &  $15.1$  & $0.37$  & $0.0030$     \\
 MR$^{+T}_{12}$-CISD($18$), $3.1609$ a$_0$  & $-4.01$ &  $20.2$  &  $-4544$ &  $17.7$  & $0.45$  & $0.0032$
\end{tabular}

\end{center}
\end{table}


However, the two corresponding interaction constants $W_d$ and $W_S$ should be considered jointly when bounds on
underlying (${\cal{CP}}$)-odd parameters are to be extracted \cite{MJung_review2014}. Interestingly, the ratio 
$W_d/W_S = 276 \left[ \frac{10^{18}}{e\, {\rm{cm}}} \right]$
found in the present work differs strongly from the respective ratios for the ThO system \cite{Denis-Fleig_ThO_JCP2016},
$W_d/W_S = 172 \left[ \frac{10^{18}}{e\, {\rm{cm}}} \right]$ and the ThF$^+$ system \cite{ThF+_NJP_2015}, 
$W_d/W_S = 176 \left[ \frac{10^{18}}{e\, {\rm{cm}}} \right]$.
With the present dominant sources the ${\cal{P,T}}$-odd energy shift is written as
\begin{equation}
 \Delta \varepsilon_{\cal{P,T}} = \frac{1}{2}\, \left( W_d\, d_e + W_c\, C_S \right)\, 
                                             \left< {\bf{n}}\cdot{\bf{e}}_z\right> (E_{\text{ext}})
 \label{EQ:PTODD_SHIFT}
\end{equation}
where ${\bf{n}}$ is the unit vector along the molecular axis and the external electric field is assumed to be
along the $z$ axis in the laboratory frame. The parameter $W_c$ is related to $W_S$ as $W_c = \frac{A}{2Z}\, W_S$,
where $A$ is the nucleon and $Z$ is the proton number \cite{PhysRevA.84.052108}. This means that an accurate
measurement of an upper bound to $\Delta \varepsilon_{\cal{P,T}} ({\text{TaO}}^+)$ in combination with the present 
values for the interaction constants $W_d$ and $W_S$ and the corresponding data for the ThO system would lead to
stronger constraints on the BSM parameters $d_e$ and $C_S$.

The nuclear MQM interaction constant $W_M = 0.45$ [$\frac{10^{33} {\text{Hz}}}{e\, {\text{cm}}^2}$] for TaO$^+$ given
in Table \ref{TAB:PTODD_PROP} is about $40$\% smaller than the one for the isoelectronic TaN system 
\cite{fleig:PRA2016}, but still sizeable. In combination with an accurate 
measurement of a ${\cal{P,T}}$-odd energy shift in the molecular ion it could be used to constrain nuclear 
(${\cal{CP}}$)-odd BSM parameters. Following the calculations in Ref. \cite{Flambaum_DeMille_Kozlov2014} the 
expected energy shift in {$^{181}$TaO$^+$ due to $|W_M\, M|$, where $M$ is the nuclear magnetic quadrupole moment, 
is $< 150\, \mu$Hz with respect to the proton EDM, $|d_p|$, as an underlying source. Correspondingly, taking the 
limits on the QCD (${\cal{CP}}$)-violating parameter $|\tilde{\Theta}|$ and the difference of the quark chromo-EDMs,
$|\tilde{d}_u-\tilde{d}_d|$, the shifts $|W_M\, M|$ correspond to $< 70\, \mu$Hz and $< 100\, \mu$Hz, respectively.
Given that the current uncertainty on the ${\cal{P,T}}$-odd energy shift of ~$800 \mu$Hz in ThO could 
ultimately be improved by two orders of magnitude \cite{ACME_ThO_eEDM_science2014,DeMille_priv}, an experiment of
similar accuracy on TaO$^+$ would probe (${\cal{CP}}$)-violating physics in the hadron sector.

The parallel component of the G-tensor in the ${^3\Delta_1}$ state of TaO$^+$, $G_{||} = 0.0032$ a.u., is very
small, even for a ${^3\Delta}$ molecule in the $\Omega = 1$ state, and has about the same size as $G_{||}$ 
in ThO (${^3\Delta_1}$) \cite{Denis_Fleig_PTmolecules_2017}; the experimental value is $g_H = 0.0044$ a.u.
\cite{kirilov_demille_ThO_PRA2013}. For comparison, $G_{||} \approx 0.03$
a.u. for ThF$^+$ (${^3\Delta_1}$) is an order of magnitude larger \cite{Denis_Fleig_PTmolecules_2017,}. The present 
finding is reasonable because there is no electronic state available in TaO$^+$
within an energy window of more than $10000$ \cm\ allowed to mix with ${^3\Delta_1}$ {\it{via}} second-order
spin-orbit interaction selection rules, see Tables \ref{TAB:TAN:SPECCON} and \ref{TAB:dipolemoments}. 
Due to this small mixing $G_{||}$ is close to $g_s-2$. This situation is qualitatively different in the
ThF$^+$ system \cite{ThF+_NJP_2015}.


\begin{table}[h]
 \caption{Molecule-frame static electric dipole moments
         $\left<{^M\Lambda}_{\Omega} | {\hat{D}}_z | {^M\Lambda}_{\Omega} \right>$, transition dipole moments
         $\left|\left|\left<{^M\Lambda}_{\Omega}' | {\hat{\vec{D}}} | {^M\Lambda}_{\Omega} \right>\right|\right|$,
         with ${\hat{\vec{D}}}$ the electric dipole moment operator (both in Debye units),
         using the model MR$^{+T}_{12}$-CISD($10$). The origin is at the center of mass, and the internuclear
         distance is $R=3.1609$ [$a_0$] (O nucleus placed at $z\vec{e}_z$ with $z>0$). Transition dipole moments
	 smaller than $10^{-7}$ Debye are not shown.
         }
 \label{TAB:dipolemoments}

 \vspace*{0.5cm}
 \hspace*{-2.8cm}
 \begin{tabular}{c|lllllll}
  ${{^M\Lambda}_{\Omega}}$ State
      & ${^3}\Delta_1$   \rule{0.4cm}{0.0cm}
      & ${^3}\Delta_2$   \rule{0.4cm}{0.0cm}
      & ${^3}\Delta_3$   \rule{0.4cm}{0.0cm}
      & \rule{0.1cm}{0.0cm} ${^1}\Sigma^+_0$ \rule{0.4cm}{0.0cm}
      & ${^3}\Sigma_0^+$ \rule{0.4cm}{0.0cm}
      & ${^3}\Sigma_1^+$ \rule{0.4cm}{0.0cm}
      & ${^1}\Delta_2$    \\ \hline
  ${^3}\Delta_1$    &\rule{0.1cm}{0.0cm} $-4.077$   & $     $  &          &          &           &          &         \\
  ${^3}\Delta_2$    &\rule{0.1cm}{0.0cm} $0.041$    & $-4.044$ & $     $  &          &           &          &         \\
  ${^3}\Delta_3$    &\rule{0.1cm}{0.0cm} $-    $    & $0.041$  & $-4.043$ & $      $ &           &          &         \\
  ${^1}\Sigma^+_0$  &\rule{0.1cm}{0.0cm} $0.036$  & $-    $  & $-    $  & $-4.109$ &           &          &         \\
  ${^3}\Sigma^+_0$  &\rule{0.1cm}{0.0cm} $0.108$  & $-    $  & $-    $  & $0.716$  & $-4.960$  &          &         \\
  ${^3}\Sigma^+_1$  &\rule{0.1cm}{0.0cm} $0.006$  & $0.087$  & $-    $  & $0.040$  & $0.000$   & $-5.341$ &         \\
  ${^1}\Delta_2$    &\rule{0.1cm}{0.0cm} $0.131$    & $0.131$  & $0.098$  & $-    $  & $-     $  & $0.071$  & $-3.309$
 \end{tabular}
\end{table}


A consistent trend is observed for the parallel magnetic hyperfine constants when comparing TaO$^+$ with three
other molecular EDM candidates. On the absolute, $A_{||}$ for {$^{229}$Th} increases by about $45$\% from ThO 
\cite{Fleig2014} to the isoelectronic ion with the heavier partner nucleus, ThF$^+$ \cite{ThF+_NJP_2015}. Likewise,
the replacement of the nitrogen by an oxygen nucleus on the neighboring atom increases $A_{||}$ for {$^{229}$Ta}
by about $54$\%. A corresponding measurement would reveal how well the spin density close to the heavy nucleus
is described by the present molecular wavefunctions which in turn would yield information on the accuracy of the
presented ${\cal{P,T}}$-odd interaction constants.

Finally, the center-of-mass molecular static electric dipole moment in TaO$^+$ (${^3\Delta_1}$) of $-4.01$ [$D$] is 
large and allows for near complete polarization of the system in weak external electric fields, for instance by
the use of ion traps and rotating electric fields \cite{Cornell_MolIons_Science2013}.

\section{Conclusion}
\label{SEC:SUMM}
In the present work a new molecular ionic system is proposed as a candidate for the
detection of an EDM signal. TaO$^+$ provides all the advantages of molecules in ${^3\Delta_1}$
states exploited in leading EDM experiments \cite{ACME_ThO_eEDM_science2014,Cornell_MolIons_Science2013}
and may be used for probing (${\cal{CP}}$)-violation beyond the standard model in both the lepton and 
the hadron sectors. The obtained ${\cal{P,T}}$-odd molecular enhancements are smaller than the ones
in the ThO and the isoelectronic TaN systems, but this disadvantage may be overcompensated by the
fact that the ${^3\Delta_1}$ state is the electronic ground state of TaO$^+$, allowing for an in
principle infinite measurement time. The only other paramagnetic molecule currently considered for EDM
experiments which has a ${^3\Delta_1}$ ground state is ThF$^+$ \cite{Gresh_ThF+_JMS2016,ThF+_NJP_2015}.
However, TaO$^+$ only has a factor of 2-3 smaller ${\cal{P,T}}$-odd constants but an order of magnitude
smaller magnetic moment in ${^3\Delta_1}$ than the latter.

In reference \cite{Flambaum_DeMille_Kozlov2014} the TaN system has been graded as ``especially promising''. 
Given the present findings, the isoelectronic TaO$^+$ system is even more so.

\begin{acknowledgments}
The {\it{Agence Nationale de la Recherche}} (ANR) through grant no.
ANR-BS04-13-0010-01, project ``EDMeDM'', is thanked for financial support.
Moreover, I wish to thank David DeMille (Yale) for an inspiring discussion.
\end{acknowledgments}

\clearpage

\bibliographystyle{unsrt}
\newcommand{\Aa}[0]{Aa}








\end{document}